\documentclass[12pt]{amsart}

\usepackage{caption}
\usepackage{subcaption}
\usepackage{amssymb,amsmath}
\usepackage{latexsym}
\usepackage{graphicx}
\usepackage{geometry}               
\usepackage{color}

\usepackage{xcolor}
\usepackage{pdfsync}
\usepackage{fancyhdr}

\usepackage[applemac]{inputenc}

\usepackage[breaklinks,colorlinks,backref]{hyperref}
\hypersetup{
    colorlinks,     
    linktoc=all, 
    linkcolor=red,  
  citecolor=hyptxt,
  urlcolor=blue}
  \hypersetup{
  citebordercolor=,
  filebordercolor=green,
  linkbordercolor=blue
}
\definecolor{hyptxt}{rgb}{0.7, 0.4, 0.9}
\usepackage{bibtopic}

\newcommand{\R}{\mathbb R}

\newcommand{\ket}[1]{|\kern.3ex#1\kern.3ex\rangle}
\newcommand{\bra}[1]{\langle\kern.3ex #1 \kern.3ex|}
\newcommand{\scalar}[2]{\langle\kern.3ex #1 \kern.3ex|\kern.3ex#2\kern.3ex\rangle}
\newcommand{\norm}[1]{\|\kern.3ex#1\kern.3ex \|}

\def\lg{\langle }
\def\rg{\rangle }

\def\vs{\varsigma}
\def\vk{\varkappa}

\def\ud{\mathrm{d}}

\def\ap{a_P}
\def\sfH{\mathsf{H}}
\def\sfh{\mathsf{h}}

\begin{document}
\title[]{Smooth Bounce in affine quantization of Bianchi I}
\author[H Bergeron, A Dapor, J-P Gazeau  and P Ma\l kiewicz]{Herv\'e Bergeron, Andrea Dapor, Jean Pierre Gazeau  and Przemys\l aw Ma\l kiewicz}
\address{ISMO, UMR 8214 CNRS, Univ Paris-Sud,  France
} \email{herve.bergeron@u-psud.fr}

\address{ APC, UMR 7164 CNRS, Univ Paris  Diderot, Sorbonne Paris Cit\'e, 75205 Paris, France}\email{gazeau@apc.univ-paris7.fr}

\address{Centro Brasileiro de Pesquisas Fisicas
22290-180 - Rio de Janeiro, RJ, Brazil }\email{gazeau@apc.univ-paris7.fr}

\address{APC, UMR 7164 CNRS, Univ Paris  Diderot, Sorbonne Paris Cit\'e, 75205 Paris, France} \email{przemyslaw.malkiewicz@apc.univ-paris7.fr}

\address{National Centre for Nuclear Research, Ho\.za 69, 00-681 Warsaw, Poland} \email{Przemyslaw.Malkiewicz@fuw.edu.pl}

\address{University of Warsaw, Ho\.za 69, 00-681 Warsaw, Poland} \email{adapor@fuw.edu.pl}

\thanks{P.M. was supported by MNISW Grant ``Mobilno\'s\'c Plus". A.D. was supported by the grant of Polish Narodowe Centrum Nauki nr
2013/09/N/ST2/04312.
The work was supported by Polonium program No. 27690XK {\it Gravity and Quantum Cosmology}.}

\date{\today}

\begin{abstract}
We present the affine coherent state quantization of the Bianchi I model. As in our previous paper on quantum theory of Friedmann models, we employ a variable associated with a perfect fluid to play a role of clock. Then we deparameterize the model. A distinctive feature, absent in isotropic models, is an extra non-holonomic constraint, which survives the deparameterization and constrains the range of physical variables. The appearance of the constraint reflects the `amplification' of singularity due to anisotropy. The quantization smoothes the extra constraint and allows quantum contracting trajectories to be smoothly transformed into expanding ones. Making use of affine coherent state we develop a semiclassical description. Figures are included to illustrate our result.
\end{abstract}

\maketitle
\tableofcontents

\section{Introduction}

In our previous paper \cite{bdgm} we proposed a quantization of spatially homogeneous and isotropic spacetime, based on coherent states for the affine group (also known as wavelets). The main result of this approach was the appearance of a quantum centrifugal potential, whose effect is the regularization of the singularity, replacing the Big Bang with a bounce both at quantum and semiclassical levels. Moreover, the true Hamiltonian is essentially self-adjoint, and hence the dynamics is unitary and unambiguous. The results obtained in this simple model of cosmological collapse encouraged us to tackle further investigations concerning more elaborate models of the Universe.   

In the present work, we study the anisotropic case of Bianchi I, allowing for more degrees of freedom in the description of spacetime. The canonical classical variables $p$ and $q$ introduced in \cite{bdgm} apply equally well to the present case, so that we can make contact with our previous result. It needs to be said that we are entering an area much less visited than the Friedmann models, and thus we should expect some new difficulties.

In the FRW case \cite{bdgm}, apart from studying the quantum true Hamiltonian, we were also interested in the properties of quantized volume and expansion rate, as they signal the singularity by vanishing and blowing up respectively. While the quantum volume turned out to be self-adjoint, the quantum expansion rate did not. Nevertheless, it was possible to give it a semiclassical meaning through its lower symbol (i.e., its expectation value on affine coherent states, opportunely peaked at classical phase space points). In the present work, in addition to these quantities, we also have to deal with shear and distortion describing the anisotropic evolution. Classically, the shear and some of the scale factors blow up at the singularity. However, the existence of minimal volume for the universe at the quantum level prevents the unbound growth of anisotropy, i.e. it smoothly evolves through the bounce. We will not provide an explicit evolution of shear and distortion at the quantum level, but confine ourselves to giving a well-behaved semiclassical Hamiltonian, which generates a well-behaved dynamics of anisotropic variables.

We should point out that, as in any Hamiltonian quantum theory of cosmology, we must choose a degree of freedom to play the role of physical time. (In the reduced phase space scheme, this is done before quantization.) Following Schutz \cite{schutz1,schutz2}, we identify this internal clock with a barotropic fluid filling the Universe. This particular choice is technically attractive because the inclusion of fluid extends the phase space with canonical variables $T$ and $p_T$, where the fluid's momentum $p_T$ is a constant of motion. Thus, it is possible to perform a reduction to the physical phase space equipped with a time-independent true Hamiltonian. This feature may be particularly useful in future analysis of non-integrable Bianchi IX model.

In Section \ref{geom} is briefly presented the geometry of the model we wish to quantize. Its canonical formulation is described in Section \ref{canframe}, where the true Hamiltonian $h$ is defined, and the non-holonomic constraint $h > 0$ is specified. The affine coherent state quantization of the true Hamiltonian is performed in Section \ref{quantH} without taking into account this constraint. In Section \ref{constraint} we discuss different possibilities to implement the latter. It appears that the best approach consists in incorporating this constraint in the classical true Hamiltonian to be quantized. To do this, we use the procedure for coherent state quantization of distributions presented in \cite{bergaz14}. The resulting quantum Hamiltonian is presented in Section \ref{quanthet} (detailed calculations can be found in Appendix \ref{quantHT}), and its lower symbol is computed in Section \ref{semicla}. This can be thought of as an effective classical Hamiltonian accounting for quantum corrections. An approximate analytic expression for this effective Hamiltonian is used in Section \ref{numres} to study the effective dynamics of fundamental variables, volume and expansion rate. Numerical analysis of the Big Bang singularity resolution is also presented. Section \ref{conclusion} concludes with a discussion of some interesting issues of our results and possible future directions.

\section{Spacetime geometry}
\label{geom}
In the present work we study the anisotropic evolution of compact (say, with torus topology $\mathbb{T}^3$) locally flat spatial sections of the spacetime equipped with the metric
\begin{equation}\label{metric}
ds^2=-N(t)^2dt^2+\sum_i a_i(t)^2(dx^i)^2\, .
\end{equation}
We fix the spatial coordinates via $\int_{\mathbb{S}^1}dx^i=1$. The Ricci scalar reads:
\begin{equation}
R=2\bigg[\sum_i\frac{1}{Na_i}\bigg(\frac{\dot{a}_i}{N}\bigg)_{,t}-\sum_{i>j}\bigg(\frac{\dot{a}_i}{Na_i}\bigg)\bigg(\frac{\dot{a}_j}{Na_j}\bigg)\bigg]\,. 
\end{equation}
Filled with perfect fluids, these spacetimes are singular as the world-lines of co-moving observers terminate within a finite proper time. Starting from a spherical chunk of space, the anisotropic evolution will deform its initial shape. Following this observation, the spacetimes are said to have cigar or barrel type singularities, with two scale factors vanishing and the other one blowing up or two scale factors reaching a finite value and the other one vanishing, respectively, at the singularity. A more detail discussion of solutions may be found in \cite{jacobs}.
\section{Canonical framework}
\label{canframe}
The Hamiltonian constraint of general relativity in the ADM variables $(q_{ij},p^{ij})$, equipped with the Poisson bracket $\left\{q_{ij}(x),p^{kl}(x^{\prime})\right\}=16\pi G\delta_{(i}^{~k}\delta_{j)}^{~l}\delta(x-x^{\prime})$, reads \cite{adm}
\begin{equation}\label{con1}
\sfH_g=\frac{1}{\kappa}\int \left[NC^0+N_iC^i\right]~\ud^3 x\, ,
\end{equation}
where
\begin{equation}\label{con2}
C^0=-\sqrt{\det(q_{ij})}\left[~^3R+\frac{1}{\det(q_{ij})}\left(\frac{1}{2}(p^k_k)^2-p^{ij}p_{ij}\right)\right],~~C^i=-2p^{ij}_{~;j}
\end{equation}
are first-class constraints and $N$, $N_i$ are Lagrange multipliers and $\kappa=16\pi G$. The truncation of the gravitational variables to the cosmological sector translates into $q_{ij}(x):= \delta_{ij}a_i^2$ and $p^{ij}(x):= \delta_{ij}p^i$, which in turn lead to $C^i\equiv 0$ and
\begin{equation} \label{ham-g}
\sfH_g=-\frac{1}{\kappa}\frac{N}{a_1a_2a_3}\bigg(\frac{1}{2}(\sum_i a_i^2p^i)^2-\sum_i (a_i^2p^i)^2\bigg),
\end{equation}
where we defined
\begin{equation}
p^1=-\frac{1}{N}\frac{(a_2a_3)\dot{}}{a_1}, \ \ \ \ \ p^2=-\frac{1}{N}\frac{(a_3a_1)\dot{}}{a_2}, \ \ \ \ \ p^3=-\frac{1}{N}\frac{(a_1a_2)\dot{}}{a_3},
\end{equation}
satisfying $\{a_i^2,p^j\}=\kappa\delta_i^{~j}$.
The Hamiltonian (\ref{ham-g}) can be brought to a simpler, diagonal form. We achieve this by Misner (canonical) parametrization of the phase space \cite{mis}: canonical pairs are given by $(\beta^0,p_0,\beta^+,p_+,\beta^-,p_-)$, where
\begin{align}
\label{betaa}
   \begin{pmatrix}
     \beta^0    \\
      \beta^+\\
      \beta^-  
\end{pmatrix} &=  \begin{pmatrix}
   1/3  &   1/3 &  1/3  \\
  1/6&   1/6 &  - 1/3\\
       \dfrac{1}{2\sqrt{3}} &  - \dfrac{1}{2\sqrt{3}} &  0
\end{pmatrix} \,\begin{pmatrix}
      \ln a_1    \\
      \ln a_2\\
      \ln a_3  
\end{pmatrix}\, ,   \\
\label{moma}  \begin{pmatrix}
      p_0    \\
      p_+\\
      p_-  
\end{pmatrix}  &=   \begin{pmatrix}
    1  &  1 & 1  \\
     1 &  1 & -2\\
     \sqrt{3} & - \sqrt{3} & 0
\end{pmatrix}\,\begin{pmatrix}
      2a_1^2\,p_1    \\
        2a_2^2\,p_2\\
        2a_3^2\,p_3
\end{pmatrix}\, . 
\end{align}
In term of these variables, the gravitational Hamiltonian reads
\begin{equation}
\sfH_g=\frac{1}{\kappa}N\frac{e^{-3\beta^0}}{24}\left(-p_0^2+p_+^2+p_-^2\right)\, . 
\end{equation}
At this point, $\sfH_g$ is still a constraint. To make it into the physical true Hamiltonian, we include matter in the form of a barotropic fluid subject to the equation of state $\mathrm{p}=w\rho$ ($w = $ const). Following Schutz \cite{schutz1,schutz2} we obtain for this matter the Hamiltonian
\begin{equation}\label{fluid1}
\sfH_m=N\frac{p_T}{(\sqrt{q})^{w}}=Ne^{-3w\beta^0}p_T\,,
\end{equation}
where $T$ and $p_T$ are canonical variables associated with the fluid. In analogy with our previous work on the FRW models \cite{bdgm}, we now replace $(\beta^0,p_0)$ with the following canonical pair:
\begin{equation}
\label{qpdef}
q:=e^{\frac{3(1-w)}{2}\beta^0},~~p:=-\frac{2}{3(1-w)}e^{-\frac{3(1-w)}{2}\beta0}p_0\, , 
\end{equation}
such that $[q]=1$, $[p]=L^2$ and $\{q,p\}=\kappa$. At the physical level we notice that this is exactly the same pair of observables which  we considered for the FRW models, i.e., 
\begin{equation}\label{int1}
q=(a_1a_2a_3)^{\frac{1-w}{2}},~~~~p=\frac{8}{3(1-w)}(a_1a_2a_3)^{\frac{1+w}{2}}\Theta\, , 
\end{equation}
where $\Theta$ is the expansion rate. The full Hamiltonian constraint reads
\begin{equation}
\sfH:= \sfH_g+\sfH_m=Nq^{\mu}\left(\frac{1}{24\kappa}\left( -\alpha^2p^2 +q^{-2}(p_+^2+p_-^2)\right) +p_T\right)\approx 0\, , 
\end{equation}
where $\mu:=-2w/(1-w)$, $\alpha:=3(1-w)/2 > 0$. Solving the constraint for $p_T$ leads to
\begin{equation}
\kappa\,  p_T=\frac{\alpha^2}{24}p^2-\frac{1}{24}(p_+^2+p_-^2)q^{-2}\, . 
\end{equation}
Upon choosing $T/\kappa$ as the clock variable, we obtain the true (non-vanishing) Hamiltonian
\begin{equation}\label{h_T}
\mathsf{h}=\frac{\alpha^2}{24}p^2-\frac{1}{24}(p_+^2+p_-^2)q^{-2}\, .
\end{equation}
describing the evolution of gravitational degrees of freedom $(q,p,\beta^-,p_-,\beta^+,p_+)\in\R_+^*\times\R^5$. The most important difference between (\ref{h_T}) and the true Hamiltonian found in \cite{bdgm} is the appearance of the attractive potential proportional to $-q^{-2}$. We must then impose positivity of $\sfh$ as a separate, non-holonomic constraint
\begin{equation}\label{pos}
\sfh>0\, ,
\end{equation}
which classically restricts the range of gravitational variables. As a side remark, we notice that in the limit of vanishing anisotropy, $p_-\rightarrow 0$ and $p_+\rightarrow 0$, the true Hamiltonian of the flat FRW spacetime \cite{bdgm} is recovered.

\section{Quantization of true Hamiltonian $\sfh$}
\label{quantH}
\subsection{Affine quantization: a compendium }
As the complex plane is viewed as the phase space for the motion
of a particle on the line, the half-plane is viewed as the phase
space for the motion of a particle on the half-line. Let  the
upper half-plane $\Pi_{+}:=\{(q,p)\,|\, p\in\mathbb{R}\,,\, q>0\}$
be equipped with the  measure $\mathrm{d}q\mathrm{d}p$. Together
with the multiplication
\begin{equation}
\label{multaff} (q,p)(q_{0},p_{0})=(qq_{0},p_{0}/q+p),\,
q\in\mathbb{R}_{+}^{\ast},\, p\in\mathbb{R}\, ,
\end{equation}
the unity $(1,0)$ and the inverse
\begin{equation}
\label{invaff} (q,p)^{-1}= \left(\frac{1}{q}, -qp \right)\, ,
\end{equation}
$\Pi_{+}$ is viewed as the affine group Aff$_{+}(\mathbb{R})$ of
the real line, and the  measure $\mathrm{d}q\mathrm{d}p$ is
left-invariant with respect to this action. The affine group
Aff$_{+}(\mathbb{R})$ has two non-equivalent unitary irreducible representations (UIR). Both are square integrable and this
is the rationale behind affine integral quantization \cite{bergaz14}. The UIR $U_{+}\equiv U$ is realized
in the Hilbert space
$\mathcal{H}=L^{2}(\mathbb{R}_{+}^{\ast},\mathrm{d}x)$:
\begin{equation}
U(q,p)\psi(x)=(e^{ipx}/\sqrt{q})\psi(x/q)\,.\label{affrep+}
\end{equation}
Picking an \emph{admissible} unit-norm state $\psi$, i.e., an element in $\in L^2(\mathbb{R}_+^\ast, \ud x)\cap L^2(\mathbb{R}_+^\ast, \ud x/x)$, 
named \emph{fiducial vector},  produces all affine  coherent states (or wavelets) defined as
\begin{equation}
\label{affwave}
| q, p \rangle\ = U(q,p) | \psi \rangle\, . 
\end{equation}
These states in $(q,p)$-representation, i.e., as functions $\zeta_{q,p}(q^{\prime},p^{\prime}):= \lg q^{\prime},p^{\prime}| q,p\rg$, are expected to be well peaked around their phase-space point $(q,p)$. 

Square integrability of the UIR $U$ and admissibility yield the resolution of the identity
\begin{equation}
\label{resunaff}
\int_{\Pi_+} \dfrac{\ud q \ud p}{2 \pi  c_{-1}} | q, p \rangle \langle q, p | = I \, , \ \mbox{where}\  
c_{\gamma}:=\int_0^\infty\frac{\ud x}{x^{2+\gamma}}\, \vert\psi(x)\vert^2\, .
\end{equation}
The covariant integral quantization of classical observables $f(q,p)$  follows: 
\begin{equation}
\label{affcovquant}
f \ \mapsto \ A_f = \int_{\Pi_+} \, f(q, p) |q, p \rangle \langle q, p | \, \dfrac{\ud q \ud p}{2\pi  c_{-1}} \, . 
\end{equation}
The above map is linear, transforms the constant function $1$ into the identity operator, and real functions into symmetric operators (provided that the integral makes sense). Also,  note that this quantization procedure can be easily extended to singular functions and distributions \cite{bergaz14}. In the sequel we restrict the choice of fiducial vector to real functions. 

The map \eqref{affcovquant} yields canonical commutation rule, up to a scaling factor,  for $q$ and $p$. Indeed, in representation ``$x$'', 
\begin{equation}
\label{affqcan}
A_p=  -i\frac{d}{d x}\equiv P\, , A_{q} =({c}_{0}/c_{-1})\, Q\, , \ Qf(x) :=  x f(x)\,, \  [A_q,A_p]=({c}_{0}/c_{-1}) iI\, . 
\end{equation}
Multiplication operator $Q$ is (essentially) self-adjoint whereas  $P$ is symmetric but  has no self-adjoint extension  \cite{reedsimon}. 
The property of the procedure which is crucial for our present purpose  is the regularization of the quantum kinetic energy:
 \begin{equation}
\label{kinquantaff}
 A_{p^2} = P^2 +   \vs_{\psi}Q^{-2}\  \mbox{with}\  \vs_{\psi}:= \int_0^{\infty}\frac{\ud u}{c_{-1}}\,u \, (\psi'(u))^2\,.
\end{equation}
 The additional term in regard with the standard canonical quantization is a centrifugal potential $\propto Q^{-2}$ whose strength depends on the fiducial vector only, and can be made as small as one wishes through an appropriate choice of $\psi$. 
If we consider the quantum dynamics of a free motion on the open half-line $\R_+^{\ast}$, it is proved  \cite{reedsimon} that  the operator $P^2= -d^2/dx^2$ alone  in $L^2(\mathbb{R}^{\ast}_+,  \ud x)$ is not essentially self-adjoint whereas the regularized operator \eqref{kinquantaff} is for $ \vs_{\psi} \geq 3/4$.  It follows that for $ \vs_{\psi} \geq 3/4$ the quantum dynamics is unitary during the entire evolution, in particular in the passage from the motion towards the origin (contracting branch) to the motion away from it (expanding branch).

\subsection{Quantization of Hamiltonian}
We quantize the $(q,p)$ variables introduced in \eqref{qpdef} according to the above affine quantization framework, while for the other variables we choose the canonical quantization one since they  respective phase space geometry is the harmless plane $\R^2$. Making use of the formulas above (see also \cite{bdgm}), the quantum version of the true Hamiltonian (\ref{h_T}) is
\begin{equation}\label{qh_T}
\sfh\mapsto A_\sfh = \frac{\alpha^2}{24}P^2+\ap^2\frac{\alpha^2}{24} \vs_{\psi}Q^{-2}-\frac{1}{24}\frac{c_{-3}}{c_{-1}} (P_+^2+P_-^2)Q^{-2}\, ,
\end{equation}
The operators $P_+$ and $P_-$ are the usual momentum operators on the real line acting in $L^2\left(\mathbb{R}^2, \ud \beta^+\ud \beta^-\right)$. Since they are self-adjoint and commute with the Hamiltonian, we can replace them with some fixed real eigenvalues, $k_+$ and $k_-$, which will remain constant during the evolution.

As in \cite{bdgm}, we choose  $\psi$  as the  function smooth in $(0,\infty)$ and rapidly decreasing at $0^+$ and at the infinity:
\begin{equation}
\label{fiducial}
\psi \equiv \psi^{\nu}(x)= \frac{1}{\sqrt{2 x \, K_0(\nu)}}e^{-\frac{\nu}{4} \left(x\xi_{12}+\frac{1}{x\xi_{12}}\right)}, \quad \textrm{with} \,\, \nu>0 \, \  \textrm{and} \,\ \xi_{12}= \dfrac{K_1(\nu)}{K_2(\nu)}\, >0. 
\end{equation}
Here and in the following, $K_r(z)$ denotes the modified Bessel functions \cite{magnus66}. Since we deal with ratios of such functions throughout the sequel, we adopt the convenient notation 
\begin{equation}
\label{xibessel}
\xi_{rs} = \xi_{rs}(\nu)=  \frac{K_r(\nu)}{K_s(\nu)}= \frac{1}{\xi_{sr}}\, .
\end{equation}
One convenient feature of such a notation is that $\xi_{rs}(\nu) \sim 1$ as $\nu \to \infty$ (a consequence of  $K_r(\nu) \sim \sqrt{\pi/(2\nu)}$). 
Hence, the quantum Hamiltonian (\ref{qh_T}) now reads
\begin{equation}\label{qh_T2}
A_\sfh = \frac{\alpha^2}{24}P^2+\ap^2\frac{\alpha^2}{24} \vs(\nu) Q^{-2}-\frac{1}{24}\xi_{21}(\nu)^2\,  (k_+^2+k_-^2) Q^{-2}\, ,
\end{equation}
where $\vs(\nu)\equiv \vs_{\psi^{\nu}}=\left(1+\nu \, \xi_{01}\right)/4$. It is convenient to introduce the auxiliary strength parameter
\begin{equation}
\label{varaux}
\vk(\nu):=\vs(\nu)- \xi_{21}(\nu)^2\, \frac{(k_+^2+k_-^2)}{\alpha^2\ap^2}
\end{equation}
so (\ref{qh_T2}) reduces to
\begin{equation}\label{A_h}
A_\sfh= \frac{\alpha^2}{24}\left(P^2+\vk(\nu)~\ap^2~Q^{-2}\right)~.
\end{equation}
The key difficulty is the additional non-holonomic positivity constraint (also present in open ($k=-1$) FRW models), $h>0$, to which the next Section is dedicated.

\section{Analysis of positivity constraint $\sfh>0$}
\label{constraint}
In what follows we consider different possible ways of implementing the positivity constraint (\ref{pos}) in the quantum theory.

\subsection{Positivity via operator modification}
The positivity requirement of the true Hamiltonian may be imposed directly on the operator $A_\sfh$ by just putting $\vk(\nu)>0$. This leads to the condition on the amount of shear, which is expressed in terms of eigenvalues $k_+$ and $k_-$, 
\begin{equation}
k_+^2+k_-^2<\alpha^2\ap^2 \, \left(\xi_{12}\right)^2\,\xi_{1-1}\, \vs\, . 
\end{equation}
The above condition reduces the space of allowed quantum states  by restricting the spectral values associated with the shear operator. More precisely, the amount of anisotropy is bounded from above by the right hand side, meaning that a highly anisotropic spacetime cannot be realized. However, there is no reason why the Universe should be so much isotropic. On the contrary, since classically the anisotropy goes as $\sim a^{-6}$ with the overall scale factor, one reasonably expects that the Universe is more and more anisotropy-dominated the closer it is to the classical singularity. We thus conclude that the above condition for the allowed amount of anisotropy in the Universe is too restrictive, and discard this approach as unphysical.

\subsection{Positivity via domain restriction}
Let us consider the eigenvalue problem for operator $A_\sfh$ in (\ref{A_h}) on $L^2(\mathbb{R}^{\ast}_+,  \ud x)$:
\begin{equation}\label{op}
-\frac{d^2\psi}{dx^2}+\frac{\vk(\nu)~\ap^2}{x^2}\psi=\lambda\psi \ .
\end{equation}
A possible way of implementing the condition (\ref{pos}) is to ensure the self-adjointness of (\ref{op}) and then constrain the Hilbert space to the positive eigenvalues, $\lambda>0$. The difficulty in this approach is to implement such a Hilbertian projection in analytical computations, since it requires full knowledge of the spectral decomposition of $A_h$. We drop this approach for technical reasons. 

\subsection{Positivity via redefinition of classical variables}
Consider the following canonical transformation:
\begin{align}
\label{qppqp}
    q&\mapsto q^{\prime}:=\frac{1}{2}\frac{pq}{p^2-\mu^2/q^2}\,,  \quad p \mapsto p^{\prime}:= p^2-\frac{\mu^2}{q^2}\,,   \\
 \label{betbet}  \beta^{\pm}&\mapsto {\beta^{\prime}}^{\pm}:=\beta^{\pm}+\frac{p_{\pm}}{2\alpha^2\mu}\ln\left(\frac{pq-\mu}{pq+\mu}\right)\, , \quad p_{\pm}\mapsto {p^{\prime}}_{\pm}:=p_{\pm}\, , 
\end{align}
where $\mu^2:=\left(p_+^2+p_-^2\right)/\alpha^2$. In these new variables, the positivity constraint (\ref{pos}) becomes  $p^{\prime}>0$. There is also another constraint, which is $\vert q^{\prime}p^{\prime}\vert >\mu/2$. We observe that this condition splits the physical phase space into two disconnected regions. If we now define $\tilde{q}^{\prime}:=q^{\prime}p^{\prime}-\mu/2$ and $\tilde{p}^{\prime}:=\ln p^{\prime}$, then $(\tilde{q}^{\prime},\tilde{p}^{\prime})\in\mathbb{R}_+^*\times\mathbb{R}$, and the true Hamiltonian reads $\sfh=e^{\tilde{p}^{\prime}}$. The advantage here is that we do not have to impose the positivity constraint quantum mechanically, since $\sfh$ in this form is automatically positive-definite. The disadvantage of this approach is that we lose connection with the variables used for FRW models \cite{bdgm}. More importantly, the expanding and contracting branches of the Universe become disconnected. For these reasons we discard also this approach.

\subsection{Positivity via redefinition of classical expression}
Another idea is to take ${\theta (\sfh)\sfh}$ (where $\theta$ is the Heaviside function) as the classical true Hamiltonian, and quantize it in the spirit of integral quantization framework \cite{bergaz14,balfrega14}. The resulting operator $A_{\theta (\sfh)\sfh}$ will have the positivity constraint implemented. Moreover, in this way we do not redefine the initial Hilbert space. In the next Section we follow this last idea, in what we call ``refined quantization'' (as opposed to Section \ref{quantH}).

\section{Refined quantization of true Hamiltonian}
\label{quanthet}
\subsection{General formula for the operator}
Let us consider the simplified form of the Hamiltonian \eqref{h_T}  
\begin{equation}
\label{h_Tbis}
\sfh_s=p^2-\frac{k^2}{q^2}
\end{equation}
obtained after dropping unnecessary factors, together with the positivity constraint $\sfh_s>0$. We derive from the decomposition $\sfh_s= \theta(\sfh_s)\sfh_s + \theta(-\sfh_s)\sfh_s$ and the linearity of the affine integral quantization the equation
\begin{equation}\label{h1}
A_{\theta(\sfh_s)\sfh_s} = A_{\sfh_s}  - A_{\theta(-\sfh_s)\sfh_s}\, .
\end{equation}
With the fiducial vector \eqref{fiducial} and from results given in \cite{bdgm}, we obtain for the affine quantized version  of $\sfh_s$ in ``$x$'' representation:
\begin{equation}
\label{quanthamT}
\lg x^{\prime}|A_{\sfh_s}|x\rg=\left[-\frac{d^2}{d x^2}+\left(\vs(\nu)- \xi_{21}(\nu)^2\, k^2\right) \dfrac{1}{x^2}\right]\delta(x-x^{\prime})\,.
\end{equation}
All the details about the derivation of this expression are given in Appendix \ref{quantHT}. 

The second term, $A_{\theta(-\sfh_s)\sfh_s}$ for the same fiducial vector (\ref{fiducial}), in ``$x$'' representation reads
\begin{align}
\lg x^{\prime}|A_{\theta(-\sfh_s)\sfh_s}|x\rg=\frac{2}{\pi K_0(\nu) c_{-1}}\left(\frac{k \mathrm{Re}[K_0(2\gamma)]}{\sqrt{xx^{\prime}}(x-x^{\prime})^2}-\frac{4 \xi_{12}}{\nu}\frac{\sqrt{xx^{\prime}}\mathrm{Im}[\gamma K_{1}(2\gamma)]}{(x-x^{\prime})^3(x+x^{\prime})}\right)\, , 
\end{align}
where
\begin{align}
\gamma=\frac{\nu}{4}\sqrt{\left(\frac{1}{x}+\frac{1}{x^{\prime}}\right)(x+x^{\prime}-i\frac{4}{\nu}\frac{k}{\xi_{12}}(x-x^{\prime}))}=\frac{\nu}{4}\sqrt{\left(1+\frac{1}{y}\right)(1+y-i\frac{4}{\nu}\frac{k}{\xi_{12}}(1-y))}
\end{align}
with $y=x^{\prime}/x$. This is a non-local operator, and there is little chance to solve its eigenvalue problem analytically. We are not going to analyze it except for noticing that the operator $A_{\theta(\sfh_s)\sfh_s}$ is positive-definite and as such it admits self-adjoint extension(s), e.g., Friedrich's extension. Therefore, the existence of unitary evolution is guaranteed. In what follows we make use of the semiclassical description available in our approach based on affine coherent states.

\section{Semiclassical approach}
\label{semicla}

\subsection{From Lagrangian}
Inspired by Klauder's approach \cite{klauderscm1,klauderscm2}, we present
a consistent framework allowing to approximate the quantum
Hamiltonian and its associated dynamics by
making use of  a semiclassical Lagrangian approach.

For a general Hamiltonian operator $\sfH$, the Schr\"odinger equation, $i \hbar \frac{\partial }{\partial t}
| \Psi(t) \rg = \mathsf{H} | \Psi(t) \rg$, can be deduced
from the Lagrangian:
\begin{equation}
\label{lagrangen} {\sf L} (\Psi, \dot \Psi, \mathcal{N}):=\lg
\Psi(t) |\left( i\hbar\frac{\partial}{\partial t}
-\mathsf{H}\right) |\Psi(t)\rg \, ,
\end{equation}
through the application of the  variational principle with respect to $|\Psi(t)\rg$. Following Klauder, we assume that  $|\Psi(t)\rg$
is an affine coherent state $| q(t),p(t)\rg$, where $q(t)$ and $p(t)$ are some time-dependent functions. Then
the Lagrangian  \eqref{lagrangen} turns
to the semiclassical form
\begin{align}
\label{lagrangian(i)sc} \nonumber \tilde{\sf L} (q,\dot q,
p) &=\lg q(t),p(t) |\left(
i\hbar\frac{\partial}{\partial t} -
\mathrm{H}
\right)|q(t),p(t)\rg \\ & = - \xi_{02}(\nu) q\dot p - \lg q(t),p(t)|
\mathrm{H} |q(t),p(t)\rg \, .
\end{align}
We rescale the family of coherent states to define a new family of coherent states $|q(t),p(t)\rg^{new}:=|\lambda q(t),p(t)\rg$, where $\lambda=\xi_{20}(\nu)$. In this way, we ensure that $(\check{q},\check{p})$, computed as the expectation values wrt to the rescaled family, fully correspond to the classical pair, i.e. $(\check{q},\check{p})=(q,p)$. Then, we derive the semiclassical equations of motion as
\begin{align}
\label{eqmot1}
  \dot q   & =  \frac{\partial}{\partial p}
 \check{\mathsf{H}}(q,p),\\ \label{eqmot2}
   \dot p &= -  \frac{\partial}{\partial q}
  \check{\mathsf{H}} (q,p) \, .
\end{align}
where $ \check{\mathsf{H}}(q,p)$ is the lower symbol of $\mathsf{H}$, $\check{\mathsf{H}}(q,p)=\lg \lambda q(t),p(t)|
\hat{\mathsf{H}} |\lambda q(t),p(t)\rg$. We note that the classical relation between $\dot{q}$ and $p$ cannot hold any longer on the semiclassical level since this relation is now given by (\ref{eqmot1}), in which classical $\mathsf{H}$ is replaced with $\check{\mathsf{H}}$. This is viewed as the consequence of the quantum non-commutativity of basic variables. In our reconstruction of  the semiclassical description of spacetime we will keep the interpretation of $q$ as fundamentally given by (\ref{int1}).

We now specify $
\mathsf{H} =
A_{\theta(\sfh_s)\sfh_s}$, and $\check{\sfH}=\lg\lambda q,p|
A_{\theta(\sfh_s)\sfh_s}|\lambda q,p \rg$. The latter  lower symbol is the basis of our semiclassical description.

\subsection{Lower symbol}

We find that the lower symbol of $A_{\theta(-\sfh_s)\sfh_s}$ reads
\begin{align} \label{h3}
\nonumber\langle\lambda q,p|A_{\theta(-\sfh_s)\sfh_s}|\lambda q,p\rangle&= \frac{2}{\pi}\,\left(\frac{4\xi_{12}}{\lambda q\,\nu\, K_0(\nu)}\right)^2\, \times \\
&\times \int_0^{\infty}\left(k\frac{\mathrm{Re}[K_{0}(2\gamma)]}{y(1-y)^2}-\frac{4\xi_{12}}{\nu}\frac{\mathrm{Im}[\gamma K_{1}(2\gamma)]}{(1-y)^3(y+1)}\right)\frac{\tilde{\gamma}^2 K_2(2\tilde{\gamma})}{(y^2+ 1)^2}\frac{y^2\,\ud y}{c_{-1}}
\end{align} 
where $$\tilde{\gamma}=\frac{\nu}{4}\sqrt{\left(1+\frac{1}{y}\right)(1+y-i\frac{4}{\nu}\frac{p\lambda q}{\xi_{12}}(1-y))}$$ and $y=x^{\prime}/x$. The analytic evaluation of the above integral is rather intractable  and we  proceed with a numerical integration instead.

\section{Singularity resolution}
\label{numres}
\subsection{Numerical examples}
\begin{figure}[t]
\begin{tabular}{cc}
\includegraphics[scale=0.3]{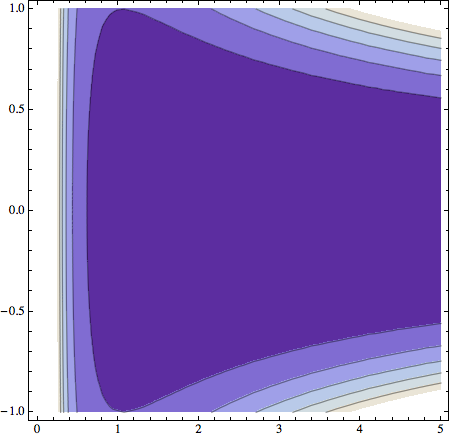}
\end{tabular}
\caption{\small The figure shows the trajectories generated by the lower symbol of the true Hamiltonian. The trajectories are clearly reversed and the bounce occurs. The parameters are chosen as  $k=10\,a_P$ (for the potential strength), $\nu=3$ (for the fiducial vector).} 
\label{figure1}
\end{figure}

\begin{figure}[t]
\begin{tabular}{ccc}
\includegraphics[scale=0.3]{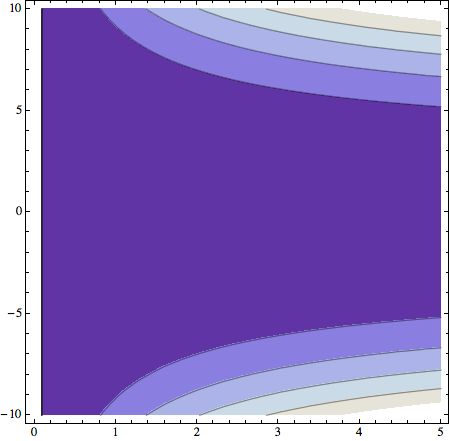} & 
\includegraphics[scale=0.3]{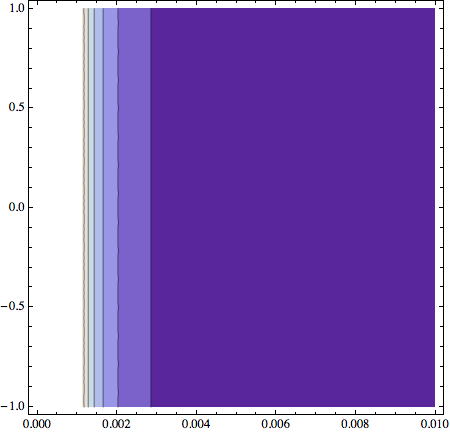}
\end{tabular}
\caption{\small Here $k=50\,a_P$, the semiclassical trajectories are initially more divergent, but they are again reversed. On the left, the same scale as above. On the right, the plot is zoomed in, so that we can see the reversal of trajectories near $q=0$.} 
\label{figure2}
\end{figure}

The semiclassical refined true Hamiltonian is defined as the difference of the lower symbol of (\ref{quanthamT}) and the lower symbol (\ref{h3}). Figures \ref{figure1} and \ref{figure2} illustrate the singularity resolution in the half-plane $(q,p)$ for different values of the strength $k$. The contour plots of the lower symbol of the Hamiltonian contain semiclassical trajectories as constant-value levels. The numerical computations prove that they exhibit bouncing behavior.

\subsection{Lower symbol of the Hamiltonian}

Combining the lower symbol of the operator $A_{\sfh_s}$, given in Appendix \ref{quantHT}, with (\ref{h3}), we find that the lower symbol of the true Hamiltonian (\ref{h1}) has the following form:
\begin{equation}
\label{lsAhs}
\lg \lambda q,p|
A_{\theta(\sfh_s)\sfh_s}|\lambda q,p \rg=\frac{1}{\lambda^2q^2}\left(p^2\lambda^2 q^2+A(\nu)-B(\nu)k^2+F_{\nu}(k^2,p^2\lambda^2q^2)\right) ,
\end{equation}
 where
\begin{align}
A(\nu):=\frac{a_P^2}{4}\,\xi_{10}\,\xi_{12}\,\left(\nu\,\xi_{32}-1\right),~ B(\nu):= \xi_{20}\,,
\end{align} 
and
\begin{align}
 \frac{1}{\lambda^2q^2}F_{\nu}(k^2,p^2\lambda^2q^2):=-\lg \lambda q,p|
A_{\theta(-\sfh_s)\sfh_s}|\lambda q,p \rg
 \end{align} 
is the corrective term due to the positivity constraint, for which we find the following limits:
\begin{equation}
F_{\nu}(0,0)=0\, ,~~\lim_{k\rightarrow\infty}F_{\nu}(k^2,0)=-A(\nu)+B(\nu)k^2\, ,~~\lim_{x\rightarrow\infty}F_{\nu}(k^2,x)=0\,.
\end{equation}
Note that the lower symbol \eqref{lsAhs} is even in $k$ and in $pq$.

\begin{figure}[t]
\begin{subfigure}[t]{.49\textwidth}
\centering
\includegraphics[width=.6\linewidth, height=.6\linewidth]{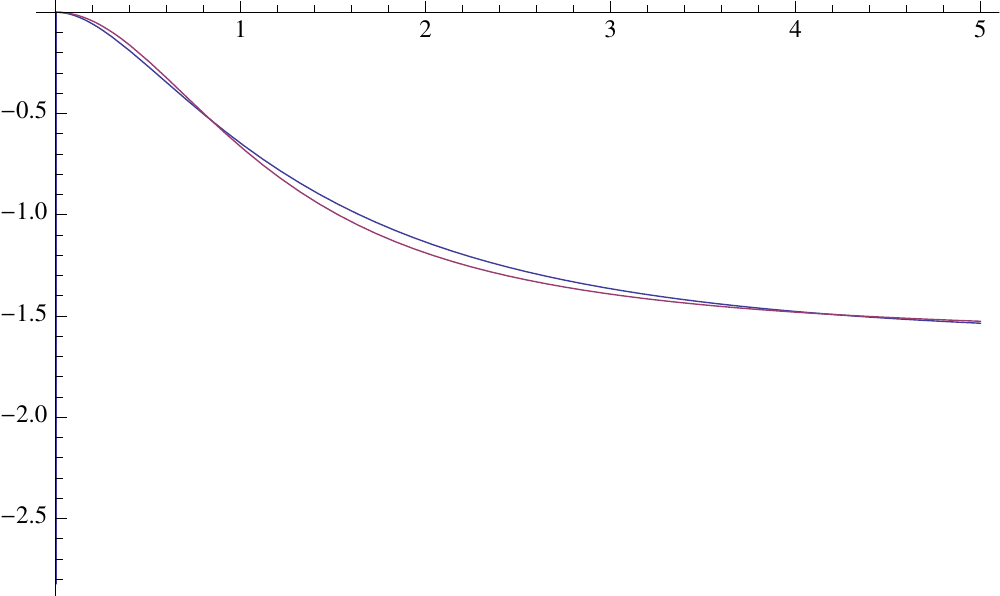}
\caption{\ref{figure4}} 
\label{figure4}
\end{subfigure}
\begin{subfigure}[t]{.49\textwidth}
\centering
\includegraphics[width=.6\linewidth, height=.6\linewidth]{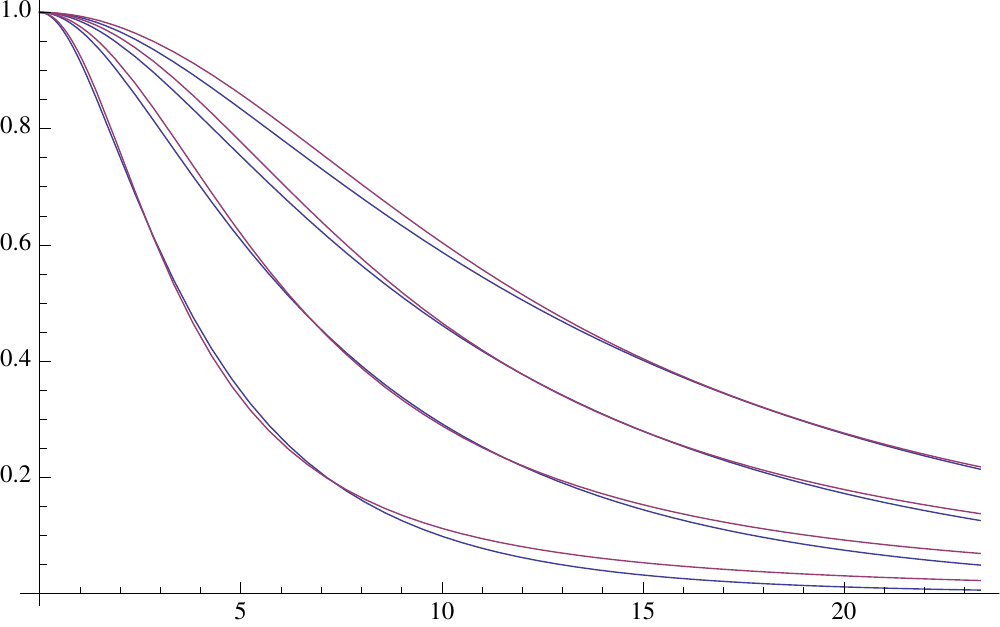}
\caption{\ref{figure3}} 
\label{figure3}
\end{subfigure}
\caption{\small (A) Plot of the exact $F_{3}(k^2,0)$ in blue and its approximation $a_{3}(k^2)$ in red. (B) Plot of the exact $\frac{F_{3}(k^2,x^2)}{F_{3}(k^2,0)}$ in blue and its approximation $b_{3}(k^2,x^2)$ in red. Plotted for $k=1,2,3,4$. Beyond the plotted range of $x$ the effect of theta function becomes negligible due to the relatively large positive value of $\lg \lambda q,p|
A_{\theta(\sfh)\sfh_s}|\lambda q,p \rg$, and the accuracy of the approximation is not very significant for reproducing the correct dynamics.} 
\end{figure}

Based on our  numerical simulations,  we can guess the following approximative form  for $F_{\nu}(k^2,p^2\lambda^2q^2)$:
\begin{equation}
F_{\nu}(k^2,p^2\lambda^2q^2)\approx a_{\nu}(k^2)\, b_{\nu}(k^2,p^2\lambda^2q^2)
\end{equation}
where
\begin{align}
&a_{\nu}(k^2)=-A(\nu)\frac{\lambda_1(\nu)k^2}{1+\lambda_1(\nu)k^2}+B(\nu)k^2,\\&b_{\nu}(k^2,p^2\lambda^2q^2)=\frac{1+\lambda_1(\nu)k^2}{1+\lambda_1(\nu)k^2+\lambda_2(\nu)p^2\lambda^2q^2}
\end{align}
That this factorization is indeed a good approximation of $F_{\nu}(k^2,p^2\lambda^2q^2)$ is confirmed for $\nu=3$ by the plots in Figures \ref{figure4} and \ref{figure3} with $\lambda_1\approx 0.3$ and $\lambda_2\approx 0.7$.

\subsection{Effective dynamics of $q$ and $p$}
Let us now restore the original physical factors $\alpha$ and $a_P$ present in \eqref{qh_T}. The approximate form of the lower symbol \eqref{lsAhs} of the true Hamiltonian now reads
\begin{equation}
\label{tildeh}
\begin{split}
\lg \lambda q,p|
A_{\theta(\sfh)\sfh}|\lambda q,p \rg&\approx \frac{1}{\lambda^2q^2}\frac{\alpha^2}{24}\left(p^2\lambda^2q^2+A(\nu)-B(\nu)\frac{k^2}{\alpha^2}+a_{\nu}\left(\frac{k^2}{\alpha^2}\right)\, b_{\nu}\left(\frac{k^2}{\alpha^2},p^2\lambda^2q^2\right)\right)\\
&\equiv \tilde \sfh(q,p)\, . 
\end{split}
\end{equation}
Making use of the approximation we obtain
\begin{equation}
(pq)\dot{}=\{pq,\tilde{\sfh}\}=2\tilde{\sfh}>0
\end{equation}
where $\tilde{\sfh}$, our semiclassical Hamiltonian, is constant and thus
\begin{equation}\label{const}
pq=2\tilde{\sfh}\, (T-T_{bounce})\, .
\end{equation}
We set $T_{bounce}=0$. Now we easily integrate (\ref{eqmot1}-\ref{eqmot2}) and obtain for $q$ the expression
\begin{equation}
q=\sqrt{\frac{q^2\tilde{\sfh}}{\tilde{\sfh}}}=\sqrt{\frac{1}{\tilde{\sfh}}\frac{\alpha^2}{24\lambda^2}\left(4\tilde{\sfh}^2T^2\lambda^2+A(\nu)-B(\nu)\frac{k^2}{\alpha^2}+a_{\nu}\left(\frac{k^2}{\alpha^2}\right)\, b_{\nu}\left(\frac{k^2}{\alpha^2},4\tilde{\sfh}^2T^2\lambda^2\right)\right)}\, . 
\end{equation}
where we substituted $pq$ according to (\ref{const}) and where $\tilde{\sfh}$ on the right-hand side is treated as a constant of integration. Next, we find for $p$
\begin{equation}
p=\frac{2\tilde{\sfh} T}{\sqrt{\frac{1}{\tilde{\sfh}}\frac{\alpha^2}{24\lambda^2}\left(4\tilde{h}^2T^2\lambda^2+A(\nu)-B(\nu)\frac{k^2}{\alpha^2}+a_{\nu}\left(\frac{k^2}{\alpha^2}\right)\, b_{\nu}\left(\frac{k^2}{\alpha^2},4\tilde{\sfh}^2T^2\lambda^2\right)\right)}}\,. 
\end{equation}

\subsection{Effective spacetime}
From the initial definitions given in Section \ref{canframe} we have
\begin{align}
ds^2=-N^2\ud T^2+(a_1dx^1)^2+(a_2dx^2)^2+(a_3dx^3)^2\, , 
\end{align}
where
\begin{align}
N=-\frac{1}{q^{\mu}},~a_1=q^{\frac{1}{\alpha}}e^{\beta^++\sqrt{3}\beta^-},~~a_2=q^{\frac{1}{\alpha}}e^{\beta^+-\sqrt{3}\beta^-},~a_3=q^{\frac{1}{\alpha}}e^{-2\beta^+}\,. 
\end{align}
In order to construct the semiclassical spacetime, in addition to $q(T)$, we need the semiclassical dynamics of $\beta^{\pm}$, which can be derived from the respective Hamilton equations after replacing $k^2$ with $p_+^2+p_-^2$ in (\ref{tildeh}). However, the resulting formulas are involved and will be omitted. Instead we focus on the overall expansion and the total volume.

\subsection{Effective dynamics of volume and expansion rate}
The volume $V$ and expansion rate $\Theta$ read respectively:
\begin{equation}
V(T)=q^{\frac{3}{\alpha}},\ \ \ \ \ \Theta(T):=\frac{1}{N}\frac{\dot{V}}{V}=-\frac{3}{\alpha}q^{\mu-1}\dot{q}
\end{equation}
Towards the singularity, the volume decreases until it reaches its minimal value, which is
\begin{equation}\label{vmin}
V_{\mathrm{min}}=\left(\frac{A(\nu)}{1+\lambda_1\frac{k^2}{\alpha^2}}\frac{\alpha^2}{24\lambda^2\tilde{\sfh}}\right)^{\frac{3}{2\alpha}}.
\end{equation}
We note that for $k^2=0$ the above formula gives the minimal volume of the flat FRW universe \cite{bdgm}. The precise moment of the maximal value of expansion rate, $T_{\Theta}$, is very difficult to obtain. Let us assume that in the vicinity of the bounce, where the contraction reaches it maximal value, we may apply the following approximation:
\begin{equation}
q\approx\sqrt{\frac{1}{\tilde{\sfh}}\frac{\alpha^2}{24\lambda^2}\left(4\tilde{\sfh}^2T^2\lambda^2+\frac{A(\nu)}{1+\lambda_1(\nu)\frac{k^2}{\alpha^2}}\right)}\, .
\end{equation}
Then, we find
\begin{equation}
T_{\Theta}\approx\pm\sqrt{\frac{1}{4\lambda^2\tilde{\sfh}^2(1-\mu)}\frac{A(\nu)}{1+\lambda_1(\nu)\frac{k^2}{\alpha^2}}}\, , 
\end{equation}
and
\begin{equation}
\Theta_{max}=\Theta(T_{\Theta})\approx\pm\frac{\alpha}{4}\left(\frac{\mu-2}{\mu-1}\frac{\alpha^2}{24\tilde{\sfh}\lambda^2}\frac{A(\nu)}{1+\lambda_1(\nu)\frac{k^2}{\alpha^2}}\right)^{\frac{\mu-1}{2}}\sqrt{\frac{1}{1-\mu}\frac{A(\nu)}{1+\lambda_1(\nu)\frac{k^2}{\alpha^2}}}\ .
\end{equation}

\section{Discussion}
\label{conclusion}

\subsection{Mechanism behind the singularity resolution}

In the present paper we have dealt with the anisotropic cosmological model of Bianchi I. Classically, such spacetime presents a singularity which is much stronger than the one for the Friedmann models considered in our previous paper \cite{bdgm}. The classical trajectories in the $(q,p)$ plane diverge as they approach the singular point. Indeed, the contracting and expanding branches are disconnected manifolds of the constraint surface. Therefore, one does not expect that canonical quantization may resolve such singularities.

\begin{figure}[t]
\includegraphics[scale=0.6]{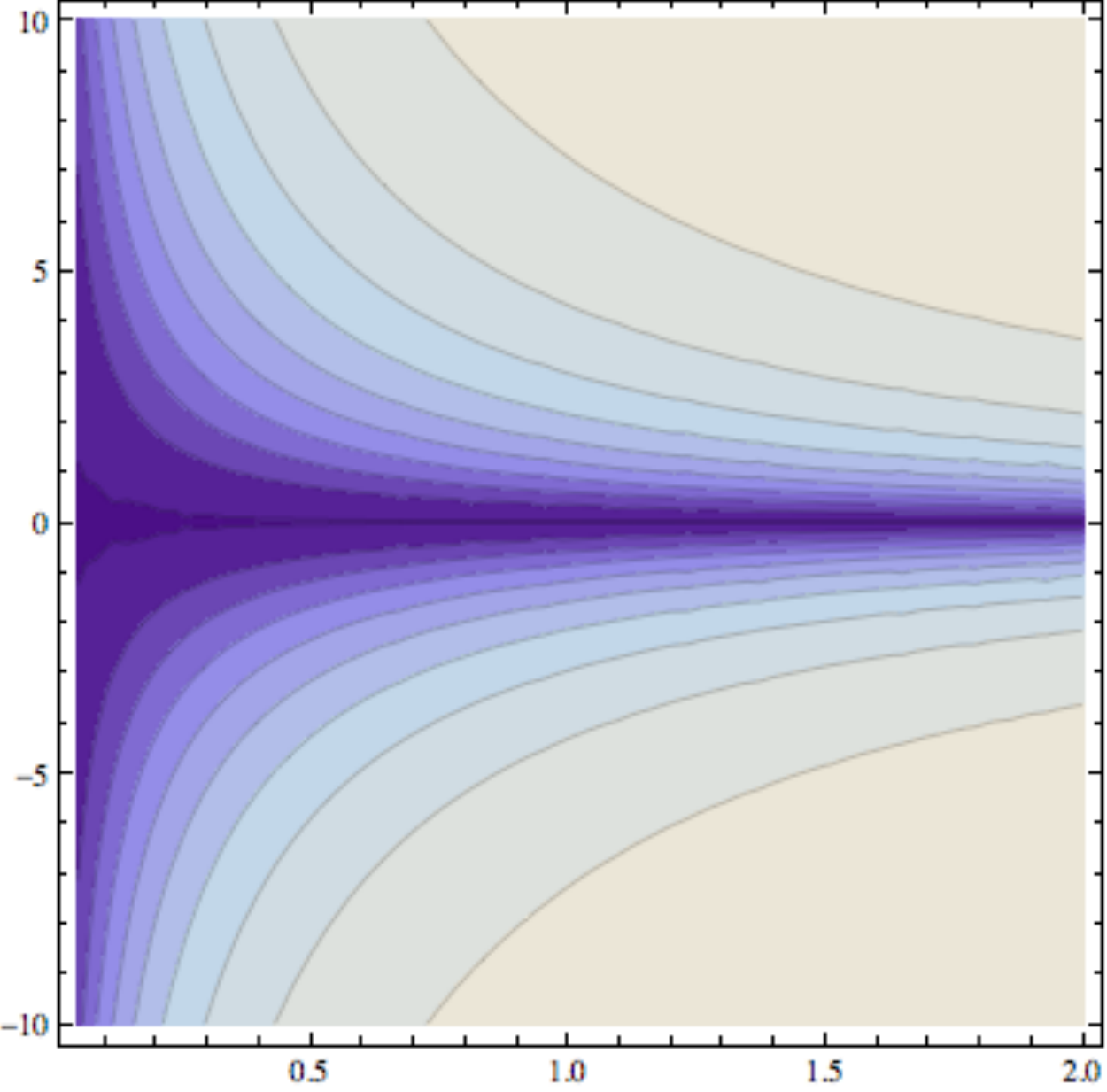} 
\caption{Contour plot of the semiclassical expression $ \lg q,  p | A_{\theta} | q , p \rg$. Here $k=3$, $q$ is running on the interval $[0,2]$ and $p$ on $[-10,10]$. $ \lg q, p | A_{\theta} | q, p \rg$ is running in the interval $[0.74,1]$.} 
\label{figureTheta}
\end{figure}

We employ a more suitable framework, known as the affine coherent state quantization (ACS). In order to resolve the singularity with a bounce, we introduce a phase space in which the expanding and contracting branches are connected by a classically forbidden region. Classically, an extra \emph{positivity constraint} is present, to ban that region. The ACS quantization, because it can also be applied to distributions, smoothes this constraint, allowing the semiclassical trajectories to cross the classically forbidden region and join the contraction and expansion with a smooth bounce (see Figure \ref{figure1}). Thus, we have obtained an anisotropic singularity resolution by a new mechanism, which is peculiar to our quantization framework. To illustrate the smoothing which takes place in the phase space, we analyze the case of theta function in Appendix \ref{aptheta}. Upon quantization, the theta function becomes a positive operator with a trivial kernel and its lower symbol is presented in Figure \ref{figureTheta}. Consequently, all quantum configurations corresponding to the phase space become accessible by the physical (i.e. satisfying all the classical and quantum constraints) motion.

The present result shows that, in order to solve the singularity problem, one needs (at least partially) to impose the gravitational constraints at the quantum level. It does not mean that the Dirac approach is preferred. Rather, a combination of reducing the constraint partially on the classical level and partially on the quantum level seems to be the best option.

\subsection{Features of the quantum bounce}
We notice in formula (\ref{vmin}) that the larger the anisotropy $k^2$, the smaller the minimal volume which is reached by the collapsing universe. This straightforward observation challenges the relevance of the Planck scale to the big bounce as the change in anisotropy may make the bounce scale arbitrarily low.\footnote
{
Another argument against the role of Planck scale in dynamics of quantum cosmological models stems from the lack of privileged time function in quantum gravity. A detailed reasoning may be found in \cite{malk2014}. Even for a specific choice of time, it has recently been shown \cite{assa2014} that quantum gravitational effects could be detected by relatively low-energy particles, provided that the spacetime is in a sufficiently non-classical state.
}
Moreover, we notice that in (\ref{vmin}) the minimal volume depends only on one effective constant $A(\nu)$, that is, the one which was present already in the isotropic models. Thus, we have obtained an extension of our previous result \cite{bdgm}, which is exactly the ``isotropic limit" of our present result. The next step in our research is to study in similar manner other types of singularity. In farther future, we will consider the effect of the obtained singularity resolution on cosmological perturbations.

\appendix

\section{Computation of quantized Hamiltonian}
\label{quantHT}
The affine CS quantization (with fiducial vector $\psi$) of the  simplified Hamiltonian $\sfh_s=p^2-k^2/q^2$ together with the positivity constraint $\sfh_s>0$ is performed through the decomposition into two terms
\begin{equation*}
A_{\theta(\sfh_s)\sfh_s} = A_{\sfh_s}  - A_{\theta(-\sfh_s)\sfh_s}\, , 
\end{equation*}
where the operators read in ``$x$''- representation, $\lg x^{\prime}|A|x\rg \equiv A$, 
\begin{align*}
  A_{\sfh_s}  &=  \int_{\Pi_+}\frac{\ud q\ud p}{2\pi}\theta\left(p^2-\frac{k^2}{q^2}\right)\left(p^2-\frac{k^2}{q^2}\right)e^{ip(x-x^{\prime})}q^{-1}\bar{\psi}(x^{\prime}/q)\psi(x/q) \, , \\
  A_{\theta(-\sfh_s)\sfh_s} &=  \int_{\Pi_+}\frac{\ud q\ud p}{2\pi}\theta\big(-p^2+\frac{k^2}{q^2}\big)\left(p^2-\frac{k^2}{q^2}\right)e^{ip(x-x^{\prime})}q^{-1}\bar{\psi}(x^{\prime}/q)\psi(x/q) \,. 
\end{align*}
First, let us focus on the second term:
\[A_{\theta(-\sfh_s)\sfh_s}=\int_{\Pi_+}\frac{\ud q\ud p}{2\pi}\theta\big(-p^2+\frac{k^2}{q^2}\big)\left(p^2-\frac{k^2}{q^2}\right)e^{ip(x-x^{\prime})}q^{-1}\bar{\psi}(x^{\prime}/q)\psi(x/q)\]
\[=\int_{\Pi_+}\frac{\ud q\ud p}{2\pi}\theta\big(|k|-|pq|\big)\left(p^2-\frac{k^2}{q^2}\right)e^{ip(x-x^{\prime})}q^{-1}\bar{\psi}(x^{\prime}/q)\psi(x/q)\]
\[=\int_{\Pi_+}\frac{\ud q\ud P}{2\pi q^3}\theta\big(|k|-|P|\big)\left(P^2-k^2\right)e^{i\frac{P}{q}(x-x^{\prime})}q^{-1}\bar{\psi}(x^{\prime}/q)\psi(x/q)\]
\[=\int_{0}^{\infty}\int_{-|k|}^{|k|}\frac{\ud q\ud P}{2\pi q^3}\left(P^2-k^2\right)e^{i\frac{P}{q}(x-x^{\prime})}q^{-1}\bar{\psi}(x^{\prime}/q)\psi(x/q)\]
\[=\int_{0}^{\infty}\int_{-|k|}^{|k|}\frac{\ud q\ud P}{2\pi q^3}\psi(x/q)\left(-q^2\frac{\partial^2}{\partial x^2}-k^2\right)e^{i\frac{P}{q}(x-x^{\prime})}q^{-1}\bar{\psi}(x^{\prime}/q)\]
\[=\int_{0}^{\infty}\frac{\ud q}{2\pi q^3}\psi(x/q)\left(-q^2\frac{\partial^2}{\partial x^2}-k^2\right)\frac{2\sin\left(\frac{|k|}{q}(x-x^{\prime})\right)}{x-x^{\prime}}\bar{\psi}(x^{\prime}/q)\]

\[=\frac{2}{\pi}\int_0^{\infty}\frac{\ud q}{q}\left(\frac{k}{q}\frac{\cos\left(\frac{|k|}{q}(x-x^{\prime})\right)}{(x-x^{\prime})^2}-\frac{\sin\left(\frac{|k|}{q}(x-x^{\prime})\right)}{(x-x^{\prime})^3}\right)\bar{\psi}(x^{\prime}/q)\psi(x/q)\,. \]
Considering now the first term,
\[A_{\sfh_s}=\int_{\Pi_+}\frac{\ud q\ud p}{2\pi}\psi(x/q)\left(-\frac{\partial^2}{\partial x^2}-\frac{k^2}{q^2}\right)e^{ip(x-x^{\prime})}q^{-1}\bar{\psi}(x^{\prime}/q)\]
\[=-\int_{0}^{\infty}\ud q\delta(x-x^{\prime})q^{-1}\bar{\psi}(x^{\prime}/q)\bigg[\psi_{,xx}(x/q)+2\psi_{,x}(x/q)\partial_x+\psi(x/q)\partial_x^2\bigg]\]
\[-k^2\int_{0}^{\infty}\ud q\delta(x-x^{\prime})q^{-3}\bar{\psi}(x^{\prime}/q)\psi(x/q)\, , \]
and introducing the integrals
\[\mathcal{I}_n:=\int_{0}^{\infty} y^n |\psi|^2\ud y~,~~\mathcal{J}_n:=\int_{0}^{\infty} y^n |\psi'|^2\ud y~,~~\mathcal{K}_n:=\int_{0}^{\infty} y^n(\bar{\psi}\psi'-\bar{\psi}'\psi)\ud y\, , \]
we get the expression
\begin{eqnarray}\nonumber
A_{\sfh_s}=\left(\mathcal{I}_{-1}P^2+\frac{i\mathcal{}\mathcal{K}_0}{2}(Q^{-1}P+PQ^{-1})+(\mathcal{J}_{1}-k^2\mathcal{I}_{1})Q^{-2}\right)\,.
\end{eqnarray}
This reduces to \eqref{quanthamT} if $\psi$ is real, since then $\mathcal{K}_0 = 0$ and we evaluate, with the notation $\xi_{rs}(\nu)= K_r(\nu)/K_s(\nu)$, 
\[\frac{\mathcal{J}_1}{\mathcal{I}_{-1}}= \vs(\nu) = \frac{1}{4}\left(1+\nu\, \xi_{01}(\nu)\right),~~~\frac{\mathcal{I}_1}{\mathcal{I}_{-1}}=\xi_{21}(\nu)^2\]
The lower symbol of $A_{\sfh_s}$ is constructed out of the following partial results:
\[
\langle q,p|P^2|q,p\rangle = p^2+\frac{1}{4}\xi_{10}(\nu)\,\xi_{12}(\nu)\,\left(\nu\,\xi_{32}(\nu)-1\right)q^{-2}
\]
and
\[\langle q,p|Q^{-2}|q,p\rangle =\xi_{10}(\nu)\,\xi_{12}(\nu)\, q^{-2}\]

\section{Quantization of $\theta$}\label{aptheta}
The kernel of $A_{\theta}$ is (dimensionless)
\begin{equation}
\lg x | A_{\theta} | x^{\prime} \rg = \delta(x-x^{\prime}) - \frac{\xi}{2 \pi K_0(\nu)} \frac{1}{x-x^{\prime}} \mathrm{Im} \left[ \sqrt{1- \frac{4 i k}{\xi \nu} \frac{x-x^{\prime}}{x+x^{\prime}}} K_1\left( \frac{\nu}{2} \frac{x+x^{\prime}}{\sqrt{x x^{\prime}}} \sqrt{1- \frac{4 i k}{\xi \nu} \frac{x-x^{\prime}}{x+x^{\prime}}} \right) \right]
\end{equation}
The semiclassical expression is
\begin{align}
 \lg q , p | A_{\theta} | q , p \rg &= \\
\nonumber 1 - \frac{\xi}{\pi K_0(\nu)^2} &  \int_0^1 \frac{\ud y }{\sqrt{y}(1-y)} \mathrm{Im} \left[  \gamma(y, k) K_1\left( \frac{\nu}{2} \frac{1+y}{\sqrt{y}}  \gamma(y, k) \right) \right] \mathrm{Re} \left[ K_0\left( \frac{\nu}{2} \frac{1+y}{\sqrt{y}}  \gamma(y, q p) \right) \right]
\end{align}
with 
\begin{equation}
\gamma(y,a) = \sqrt{1- \frac{4 i a}{\xi \nu} \frac{1-y}{1+y}}
\end{equation}


\begin{thebibliography}{99}

\bibitem{bdgm}
H Bergeron, A Dapor, J-P Gazeau and P Ma\l kiewicz, ``Smooth big bounce from affine quantization'', Phys. Rev. D \textbf{89} (2014) 083522 [arXiv:1305.0653].

\bibitem{schutz1} B. F. Schutz, ``Perfect Fluids in General Relativity: Velocity Potentials and a Variational Principle'', Phys. Rev. D \textbf{2} (1970) 2762.

\bibitem{schutz2} B. F. Schutz, ``Hamiltonian Theory of a Relativistic Perfect Fluid'', Phys. Rev. D \textbf{4} (1971) 3559.

\bibitem{bergaz14} H. Bergeron and J.-P. Gazeau, ``Integral quantizations with two basic examples'', Annals of Physics  \textbf{344} (2014) 43 [arXiv:1308.2348]. 

\bibitem{jacobs} K. C. Jacobs, ``Spatially Homogeneous and Euclidean Cosmological Models with Shear", ApJ 153 (1968).


\bibitem{adm} R. Wald, ``General Relativity'', The University Chicago Press (1984).

\bibitem{mis} C. W. Misner, in ``Magic Without Magic'', Klauder, J. (ed.), Freeman, San Francisco (1972).

\bibitem{reedsimon} M. Reed and B. Simon, ``Methods of Mathematical Physics, Vol 2'',  Academic Press (1975).

\bibitem{magnus66} W. Magnus, F. Oberhettinger and Raj~Pal Soni, ``Formulas and Theorems for the Special Functions of Mathematical Physics'', Springer-Verlag, Berlin, Heidelberg and New York (1966).

\bibitem{balfrega14} M. Baldiotti, R. Fresneda and J. P. Gazeau, ``About Dirac\&Dirac constraint quantizations'',  to appear in \textit{Physica Scripta} as an Invited Comment (2015). 

\bibitem{klauderscm1} J. R. Klauder, ``Enhanced Quantization: A Primer", J. Phys. A: Math. Theor. \textbf{45} (2012) 285304--1--8 [arXiv:1204.2870].

\bibitem{klauderscm2} J. R. Klauder, ``Completing Canonical Quantization, and Its Role in Nontrivial Scalar Field Quantization'', [arXiv:1308.4658].

\bibitem{malk2014} P. Malkiewicz, ``Multiple choices of time in quantum cosmology", [arXiv: 1407.3457].

\bibitem{assa2014} M. Assanioussi, A. Dapor, J. Lewandowski, ``Rainbow metric from quantum gravity", [arXiv: 1412.6000].

\end{thebibliography}
\end{document}